\documentstyle[preprint,aps]{revtex}
\begin{document}
\draft
\def \a1{{\alpha_1}}
\def \a2{{\alpha_2}}
\def \v{{\vec u}}
\preprint{MRI-PHY/980755}
\title{ Dissipative Phase Fluctuations In A Superconductor\\
In Proximity To An Electron Gas}
\author{D.M. Gaitonde}
\address{Mehta Research Institute,\\ 
          Chatnaag Road, Jhusi,\\
             Allahabad 211019, INDIA. \\ }
\maketitle
\begin{abstract}
We study a two-dimensional superconductor in close proximity
to a  two-dimensional metallic sheet. The electrons in the superconducting sheet
are coupled to those in the metallic sheet by the Coulomb interaction only.
We obtain an effective phase-only action for the superconductor
by integrating out all the electronic degrees of freedom in the
problem. The Coulomb drag of the normal electrons in the 
metallic sheet is found to make the spectrum of phase-fluctuations
in the superconductor, dissipative at long wavelengths.
The dissipative co-efficient $\eta$ is shown to be simply related
to the normal state conductivities of the superconducting layer
($\sigma_S$)and the metallic sheet ($\sigma_E$) by the relation
$\eta \propto {\sigma_S\sigma_E\over \sigma_S+\sigma_E}$.
\end{abstract}
\newpage

The study of phase fluctuations in disordered superconductors
and Josephson junction arrays has received a great deal
of attention [1,2,3]
over the years as a mechanism for the superconductor-insulator transition
[4,5] in systems with mesoscale disorder.
The inhomogenities in these systems exist on a length scale
much larger than the superconducting coherence length.
The pairing mechanism is therefore believed to be relatively
unaffected and the superconductor-insulator transition
seen at low temperatures with increasing disorder
is believed to be caused by enhanced fluctuations 
in the phase of the superconducting order parameter.
A striking feature of this transition is the connection
between an increasing normal state resistance 
and enhanced phase fluctuations in the superconductor
which eventually destroy global phase coherence,
thus leading to an insulating state.
The discovery of a pseudo-gap phase [6,7,8] in the underdoped
cuprates has given a renewed impetus to this field
because of the possibility [9,10] that this phase corresponds
to a collection of Cooper pairs that have lost global phase
coherence.

Recently a superconductor-insulator transition was seen
by A. J. Rimberg et. al. [11] in a two-dimensional array
of Josephson junctions held within 100 nm of a two-dimensional
electron gas (2DEG). A novel aspect of their work
was the fact that the transition was achieved by
tuning the conductivity of the 2DEG without altering the
Josephson junction array directly.
Motivated by this experiment we study, in this paper,
the phase fluctuations in a 2-D superconductor held
in close proximity (at a distance d) from a 2DEG.
We model the superconductor by a two-dimensional
BCS model together with a random potential
and the 2DEG as an electron gas with a random potential.
We explicitly include the full long-range Coulomb interaction,
both inplane and interplanar.
The Coulomb drag caused by the 2DEG is shown to make
the long wavelength phase fluctuations in the superconductor
dissipative in nature.

The dynamics of the system of electrons is described by the
action
$$S=\int_0^\beta d\tau\int d^2x\int dz [L^{sc}+L^{eg}+L^{ef}]
\eqno{(1a)}$$
where
$$L^{sc}=\delta(z-d)[\sum_{\sigma}\bar{\psi}_{\sigma}({\bf x},\tau)
({\partial\over \partial\tau}+h_s)\psi_{\sigma}
+{\mid\Delta({\bf x},\tau)\mid^2\over g}+(\Delta({\bf x},\tau)
\bar{\psi}_{\uparrow}({\bf x},\tau)\bar{\psi}_{\downarrow}
({\bf x},\tau)+h.c.)]\eqno{(1b)},$$
$$L^{eg}=\delta(z)[\sum_{\sigma}\bar{\chi}_{\sigma}({\bf x},\tau)
({\partial\over \partial\tau}+h_e)\chi_{\sigma}({\bf {x}},\tau)]\eqno{(1c)},$$
and
$$L^{ef}={(\nabla A_0({\bf x},z,\tau))^2\over 8\pi}\eqno{(1d)}.$$
The electrons at $({\bf x},\tau)$ with spin $\sigma$
are represented by the Grassman field variables
$\bar{\psi}_{\sigma}({\bf x},\tau)$, 
${\psi}_{\sigma}({\bf x},\tau)$ and
$\bar{\chi}_{\sigma}({\bf x},\tau)$,
${\chi}_{\sigma}({\bf x},\tau)$ in the 
superconducting layer (at $z=d$) and the 2DEG
(at $z=0$) respectively.
Here
$$h_s={-\hbar^2\nabla^2\over 2m_1 }-ieA_0({\bf x},d,\tau)+V_s(
{\bf x})-\epsilon_F^s \eqno{(2a)}$$
and
$$h_e={-\hbar^2\nabla^2\over 2m_2 }-ieA_0({\bf x},0,\tau)+V_e(
{\bf x})-\epsilon_F^e \eqno{(2b)}.$$
Thus,  $L^{sc}$ includes the electronic kinetic energy
and the coupling of the superconducting electrons at $z=d$
to the Coulomb potential ($A_0$) as well as a random potential
($V_s$). The field $\Delta$ is the auxilliary Hubbard-Stratonovich
field obtained from the BCS contact interaction and $g$
is the strength of the attractive interaction.
$L^{eg}$ describes the 2DEG at $z=0$ together with its coupling
to a random potential ($V_e$) and the Coulomb potential.
$L^{ef}$ gives the electric field energy of the system.
We do not consider transverse vortex-like fluctuations and so the vector
potential can be set equal to zero.

At low temperatures we can ignore fluctuations in the amplitude
of the superconducting order parameter and make the replacement[12,13]
$\Delta({\bf x},\tau)= \Delta_0
\exp{[i\theta({\bf x},\tau)]}$ where $\Delta_0$ is the mean field
value of $\mid\Delta({\bf x},\tau)\mid$.
Then on going over to a gauge in which the order parameter is real [13],
(i.e. making the transformation $\psi_{\sigma}({\bf x},\tau)
\rightarrow \exp{[{i\theta\over 2}]}\psi_{\sigma}({\bf x},\tau))$
we find that $L^{sc}$ becomes
$$L^{sc}=\delta(z-d)[L^1+L^2]\eqno{(3a)}$$
where
$$L^1=\sum_{\sigma}\bar{\psi}_{\sigma}[{\partial\over \partial\tau}+
{i\over 2}({\dot \theta}-2eA_0)+{({\hbar\nabla\over i}+{\hbar\nabla\theta
\over 2})^2\over 2m_1}+V_s({\bf x})-\epsilon_F^s]\psi_{\sigma}
\eqno{(3b)}$$
and
$$L^2={\Delta_0^2\over g}+\Delta_0(\bar{\psi}_{\uparrow}\bar{\psi}_
{\downarrow}+h.c.)\eqno{(3c)}$$

We now proceed to integrate out the fermions.
We first consider $L^{eg}$. Then we have
$$\int D\bar{\chi}D\chi e^{-S^{eg}}=e^{-S_{eff}^{eg}[A_0]}\eqno{(4a)}$$
where 
$$S^{eg}=\int_0^{\beta}d\tau\int d^3r L^{eg}\eqno{(4b)}$$.
We then compute $S_{eff}^{eg}[A_0]$ to second order in $A_0$
by performing a functional Taylor expansion of $ S_{eff}^{eg}$.
There is no contribution to $S_{eff}^{eg}$ at first order
in $A_0$ as the contribution coming from the electrons
is exactly cancelled by the positive ionic background
whose conribution has not been explicitly written earlier.
It contributes a term of the type $ieA_0({\bf x},0,\tau)
\bar{n}_{eg}$ where $\bar{n}_{eg}$ is the average electronic 
density in the 2DEG.

At second order we find that $S_{eff}$ is given by
$$S_{eff}^{eg}={e^2\over 2}\int_0^{\beta}d\tau
\int_0^{\beta}d\tau^{\prime}\int d^2x \int d^2x^{\prime}
A_0({\bf x},0,\tau) A_0({\bf x^{\prime}},0,\tau^{\prime})
P^0_{eg}({\bf x}-{\bf x^{\prime}},\tau-\tau^{\prime})
\eqno{(5)}.$$
Here 
$$P^0_{eg}({\bf x}-{\bf x^{\prime}},\tau-\tau^{\prime})
=\langle T_{\tau}[ n_{eg}({\bf x},\tau) n_{eg}({\bf x}^{\prime}
,\tau^{\prime})]\rangle \eqno {(6)},$$
 $ n_{eg}({\bf x},\tau)$ is the electron density
fluctuation operator and averages are performed with respect to 
$S^{eg}$ after setting $A_0$ to be zero.

We now integrate out the superconducting electrons.
Once again we write  
$$\int D\bar{\psi}D\psi e^{-S^{sc}}=e^{-S_{eff}^{sc}[\theta,A_0]}
\eqno{(7)}.$$
Proceeding as before,
we find at first order there is no contribution
to $S_{eff}^{sc}$.
At second order we find
$$S_{eff}^{sc}={1\over 8}\int_0^{\beta}d\tau
d\tau^{\prime}\int d^2x  d^2x^{\prime}
[P^0_{sc}({\bf x}-{\bf x^{\prime}},\tau-\tau^{\prime})
f({\bf x},\tau)f({\bf x^{\prime}},\tau^{\prime})
+D({\bf x}-{\bf x^{\prime}},\tau-\tau^{\prime})
\nabla\theta({\bf x},\tau)
\cdot\nabla^{\prime}\theta
({\bf x^{\prime}},\tau^{\prime})]\eqno{(8)}.$$
Here 
$$f({\bf x},\tau)=\dot\theta({\bf x},\tau)-
2eA_0({\bf x},d,\tau) \eqno{(9a)},$$
$$P^0_{sc}({\bf x}-{\bf x^{\prime}},\tau-\tau^{\prime})
=\langle T_{\tau}[ n_{sc}({\bf x},\tau) n_{sc}({\bf x}^{\prime}
,\tau^{\prime})]\rangle \eqno {(9b)},$$
and
$$D({\bf x}-{\bf x^{\prime}},\tau-\tau^{\prime})
={{\bar n}_{sc}\over m_1}\delta^2({\bf x}-{\bf x^{\prime}})
\delta(\tau-\tau^{\prime})-
{1\over m_1^2}\langle T_{\tau} [p_{x_1}
({\bf x},\tau) p_{x_1}({\bf x}^{\prime}
,\tau^{\prime})]\rangle \eqno {(9c)}.$$
$n_{sc}$ represents the electron density fluctuation 
operator in the superconducting state and $p_{x_1}$
is the momentum density operator along the $x_1$ direction.
All averages are performed with respect to 
$L_{sc}$ with $A_0$ and $\theta$ set equal to zero.

Including the electric field energy in $L^{ef}$
we find that the effective action for the system
becomes
$$S_{eff}=S_{eff}^{eg}+S_{eff}^{sc}+S^{ef}\eqno{(10a)}$$
where $S_{eff}^{eg}$ and $S_{eff}^{sc}$ are
defined in Eqs. (5) and (8) respectively
and 
$$S^{ef}=\int_0^{\beta}d\tau {[\nabla A_0({\bf x},z,\tau)]^2
\over 8\pi}\eqno{(10b)}$$ 

Varying the effective action $S_{eff}$ with respect to
$A_0({\bf x},z,\tau)$ we obtain the equation
of motion to be
$${\nabla^2A_0\over 4\pi}=\delta(z-d)X({\bf x},\tau)
        +\delta(z)Y({\bf x},\tau)\eqno{(11a)}$$
        where
$$X({\bf x},\tau)={-e\over 2}\int_0^{\beta}d\tau^{\prime}\int d^2x^{\prime}        
P^0_{sc}({\bf x}-{\bf x^{\prime}},\tau-\tau^{\prime})
({\partial\theta\over \partial\tau^{\prime}}-2eA_0({\bf x^{\prime}},
d,\tau^{\prime}))\eqno{(11b)}$$
and
$$Y({\bf x},\tau)=e^2\int_0^{\beta}d\tau^{\prime}\int d^2x^{\prime}        
 P^0_{eg}({\bf x}-{\bf x^{\prime}},\tau-\tau^{\prime})
A_0({\bf x^{\prime}},
0,\tau^{\prime})\eqno{(11c)}$$

Going over to Fourier space by using the transformations
$$\theta({\bf x},{\tau})={1\over \beta}\sum_{\nu_m}
\int {d^2q\over (2\pi)^2}\exp{i[{\bf q}\cdot {\bf x}-\nu_m\tau]}
\theta({\bf q},\nu_m)\eqno{(12a)}$$
and 
$$A_0({\bf x},z,\tau)={1\over \beta}\sum_{\nu_m}
\int {d^2q\over (2\pi)^2}\int {dk\over 2\pi}
\exp{i[{\bf q}\cdot {\bf x}+kz-\nu_m\tau]}
{\tilde A}_0({\bf q},k,\nu_m)\eqno{(12b)}$$
we find that Eq. (11) can be rewritten as
$${\tilde A}_0({\bf q},k,\nu_m)={1\over q^2+k^2}[-2\pi e\exp{[-ikd]}
X_1({\bf q},\nu_m)-4\pi e^2\exp{[-ikd]}X_2({\bf q},\nu_m)-
4\pi e^2X_3({\bf q},\nu_m)]\eqno{(13a)}.$$
Here
$$X_1({\bf q},\nu_m)=P^0_{sc}({\bf q},\nu_m)i\nu_m\theta({\bf q},\nu_m)
\eqno{(13b)},$$
$$X_2({\bf q},\nu_m)=P^0_{sc}({\bf q},\nu_m)A_0({\bf q},z=d,\nu_m)
\eqno{(13c)},$$
and
$$X_3({\bf q},\nu_m)=P^0_{eg}({\bf q},\nu_m)A_0({\bf q},z=0,\nu_m)
\eqno{(13d)}.$$
Eq. (13) is an integral equation for ${\tilde A}_0({\bf q},k,\nu_m)$.
Solving for $A_0({\bf q},d,\nu_m)$ and
$A_0({\bf q},0,\nu_m)$ we find, after some algebra, the relations
$$A_0({\bf q},d,\nu_m)[1+{2\pi e^2\over q}P^0_{sc}({\bf q},\nu_m)]
+{2\pi e^2\over q}e^{-qd}P^0_{eg}({\bf q},\nu_m)A_0({\bf q},0,\nu_m)
={-\pi e i\nu_m\over q}P^0_{sc}({\bf q},\nu_m)\theta({\bf q},\nu_m)
\eqno{(14a)}$$
and
$$A_0({\bf q},0,\nu_m)[1+{2\pi e^2\over q}P^0_{eg}({\bf q},\nu_m)]
+{2\pi e^2\over q}e^{-qd}P^0_{sc}({\bf q},\nu_m)A_0({\bf q},d,\nu_m)
={-\pi e i\nu_m\over q}e^{-qd}P^0_{sc}({\bf q},\nu_m)\theta({\bf q},\nu_m)
\eqno{(14b)}.$$

Solving Eqs. (14a) and (14b) for $A_0({\bf q},d,\nu_m)$ we find that
$$A_0({\bf q},d,\nu_m)={2\pi e i\nu_m \theta({\bf q},\nu_m)R_1
\over 1+R_2}\eqno{(15a)}$$
where
$$R_1={-P^0_{sc}({\bf q},\nu_m)\over 2q}-{\pi e^2\over q^2}
P^0_{sc}({\bf q},\nu_m)P^0_{eg}({\bf q},\nu_m)(1-e^{-2qd})
\eqno{(15b)}$$
and
$$R_2={2\pi e^2\over q}(P^0_{sc}({\bf q},\nu_m)+P^0_{eg}({\bf q},\nu_m))
+{4\pi^2e^4\over q^2}P^0_{sc}({\bf q},\nu_m)P^0_{eg}({\bf q},\nu_m)
(1-e^{-2qd})\eqno{(15c)}.$$

Substituting the equation of motion (Eq. (11) ) in 
the expression for $S_{eff}$ (Eq. (10)) we find
that $S_{eff}$ becomes 
$$S_{eff}=S_{eff}^1+S_{eff}^2 \eqno{(16a)}$$
where 
$$S_{eff}^1={1\over 8}\int_0^{\beta}d\tau
d\tau^{\prime}\int d^2x  d^2x^{\prime}
P^0_{sc}({\bf x}-{\bf x^{\prime}},\tau-\tau^{\prime})
{\partial\theta\over\partial\tau}
f({\bf x^{\prime}},\tau^{\prime})\eqno{(16b)}$$
and
$$S_{eff}^2={1\over 8}\int_0^{\beta}d\tau
d\tau^{\prime}\int d^2x  d^2x^{\prime}
D({\bf x}-{\bf x^{\prime}},\tau-\tau^{\prime})
\nabla\theta({\bf x},\tau)
\cdot\nabla^{\prime}\theta({\bf x^{\prime}},\tau^{\prime})
\eqno{(16c)}.$$

For simplicity, we first consider the limits
$d\rightarrow \infty$ and $d\rightarrow 0$.
In the former case we find that
$$A_0({\bf q},d,\nu_m)={-(\pi e/q)P^0_{sc}({\bf q},\nu_m)i\nu_m
\theta({\bf q},\nu_m)\over 1+(2\pi e^2/q)P^0_{sc}({\bf q},\nu_m)}
\eqno{(17)}.$$
Notice that in this limit the layers become decoupled
and $P^0_{eg}$ drops out of the expression for
$A_0$ in the superconducting layer.
Substituting Eq. (17) in the earlier relation
(Eq. (16b) ) for $S_{eff}^1$ we get
$$S_{eff}^1={1\over 8\beta}\sum_{\nu_m}\int {d^2q\over
(2\pi)^2} {\nu_m^2 \mid \theta({\bf q},\nu_m)\mid^2
P^0_{sc}({\bf q},\nu_m)\over 
1+(2\pi e^2/q)P^0_{sc}({\bf q},\nu_m)}\eqno{(18)}$$
This in combination with $S_{eff}^2$ (Eq. (16c))
is the usual action [13] for phase fluctuations in a 
two-dimensional superconductor and at long wavelengths
the phase fluctuations obey a dispersion relation
proportional to $q^{1/2}$.

Assuming that the onset of superfluid order
doesn't affect the electronic compressibility
very much, we have the relation
$$1+(2\pi e^2/q)P^0_{sc}({\bf q},\nu_m)\approx 
\epsilon_s({\bf q},\nu_m)\eqno{(19)}$$
where 
$\epsilon_s({\bf q},\nu_m)$ is the dielectric function 
of the superconductor in its normal state.
This dielectric function, in turn, is related
to the normal state conductivity ($\sigma_S$)
of the superconductor by the relation
$$\epsilon_s({\bf q},\nu_m)=1+{4\pi\hbar\sigma_S({\bf q},\nu_m)
\over \mid\nu_m\mid}\approx {4\pi\hbar\sigma_S({\bf q},\nu_m)
\over \mid\nu_m\mid}
\eqno{(20a)}$$
at low frequencies.
In a similar manner, at low frequencies we have the relation
$$P^0_{sc}({\bf q},\nu_m)\approx 
{4\pi\hbar\sigma_S({\bf q},\nu_m)
\over (2\pi e^2/q)\mid\nu_m\mid}\eqno{(20b)}.$$
Making these substitutions in Eq. (18) we find that
$S_{eff}$ reduces to 
$$S_{eff}={1\over 8\beta}\sum_{\nu_m}
\int {d^2q\over (2\pi)^2}
\mid\theta({\bf q},\nu_m)\mid^2[{q\nu_m^2\over 2\pi e^2}
+D({\bf q},\nu_m)q^2]
\eqno{(21)}$$
which is the plasma action for a two-dimensional
superconductor.

We now turn our attention to the other limit
of $d\rightarrow 0$. In this case, we find that
$$A_0({\bf q},d,\nu_m)={-\pi (e/q)i\nu_m P^0_{sc}({\bf q},\nu_m)
\over 1+2\pi e^2/q(P^0_{sc}({\bf q},\nu_m)+P^0_{eg}({\bf q},\nu_m))}
\eqno{(22)}.$$
Using the result of Eq. (22) for $A_0$ and taking the low frequency
limit as before we find that $S_{eff}$ becomes
$$S_{eff}={1\over 8\beta}\sum_{\nu_m}\int {d^2q\over (2\pi)^2}
\mid\theta({\bf q},\nu_m)\mid^2[{\hbar\sigma_S\sigma_Eq\mid\nu_m\mid
\over \sigma_S+\sigma_E}+D({\bf q},\nu_m)q^2]
\eqno{(23)}$$
where $\sigma_E$ is the conductivity of the 2DEG
and we have replaced the wavevector and frequency 
dependent conductivities by their d.c. values in the
low energy limit.
In writing Eq. (23) we have made use of relations
similar to Eqs. (19), (20a) and (20b) for the electron gas 
as well.

It is clear from Eq. (23) that in this case the phase
fluctuations are dissipative in nature and the
corresponding viscosity co-efficient ($\eta$)
is related to the conductivities of the two types of electrons
by the relation
$$\eta \propto {\sigma_S\sigma_E\over \sigma_S+\sigma_E}
\eqno{(24)}.$$

We will now consider the general expression for
$A_0({\bf q},d,\nu_m)$ (Eq. (15)).
In this case we find that $S_{eff}$ becomes
$$S_{eff}={1\over 8\beta}\sum_{\nu_m}\int {d^2q\over (2\pi)^2}
\mid\theta({\bf q},\nu_m)\mid^2[{Y_1({\bf q},\nu_m)
\over Y_2({\bf q},\nu_m)}\nu_m^2
+D({\bf q},\nu_m)q^2]\eqno{(25a)}$$
where
$$Y_1({\bf q},\nu_m)=P^0_{sc}({\bf q},\nu_m)\epsilon_{eg}
({\bf q},\nu_m)\eqno{(25b)}$$
and
$$Y_2({\bf q},\nu_m)=\epsilon_{eg}
({\bf q},\nu_m)+\epsilon_{sc}
({\bf q},\nu_m)-1+(2\pi e^2/q)P^0_{sc}({\bf q},\nu_m)P^0_{eg}({\bf
q},\nu_m) (1-e^{-2qd})\eqno{(25c)}.$$
Here $\epsilon_{eg}
({\bf q},\nu_m)=1+{2\pi e^2\over q}P^0_{eg}({\bf
q},\nu_m)$ is the dielectric function of the 2DEG.

It is clear from Eq. (25) that $d^{-1}$ defines 
a crossover scale and phase fluctuations with
$q\gg d^{-1}$ ($e^{-qd}\rightarrow 0$) behave 
like the $d\rightarrow \infty$ limit considered
earlier i.e. they are propogating modes.
In this case $Z({\bf q},\nu_m)={Y_1({\bf q},\nu_m)
\over Y_2({\bf q},\nu_m)}$ reduces to
${P^0_{sc}({\bf q},\nu_m)\over
 1+{2\pi e^2\over q}P^0_{sc}({\bf q},\nu_m)}$.
 The action for these modes becomes independent of
 $P^0_{eg}({\bf q},\nu_m)$ showing that the short wavelength modes
 are unaffected by the presence of the metallic layer.
 On further taking the low frequency limit, as before,
 $Z({\bf q},\nu_m)$ becomes ${q\over 2\pi e^2}$
 as in the special case of $d\rightarrow \infty$
 considered earlier.
 
The opposite limit of $q\ll d^{-1}$ is like the case of
$d\rightarrow 0$ treated earlier.
In that case $e^{-2qd}\rightarrow 1$
and $Z({\bf q},\nu_m)$ reduces to 
$$Z({\bf q},\nu_m)={ P^0_{sc}({\bf q},\nu_m)
\epsilon_{eg}
({\bf q},\nu_m)\over \epsilon_{eg}
({\bf q},\nu_m)+\epsilon_{sc}
({\bf q},\nu_m)-1}\eqno{(26)}.$$
Now taking the low frequency limit we find that
$Z({\bf q},\nu_m)$ becomes
$$Z({\bf q},\nu_m)={\hbar\sigma_S({\bf q},\nu_m)
\sigma_E({\bf q},\nu_m)\over \mid \nu_m \mid
[\sigma_S({\bf q},\nu_m)+\sigma_E({\bf q},\nu_m)]}
\eqno{(27)}.$$
Further in the low energy limit we can replace
the conductivities $\sigma_S({\bf q},\nu_m)$
and $\sigma_E({\bf q},\nu_m)$ by their d.c.
values $\sigma_S$ and $\sigma_E$.
Then it is clear from Eqs. (25) and (27)
that the long wavelength modes are dissipative in nature.
Their behaviour is identical with the $d\rightarrow 0$
limit considered earlier.
Notice once again that the co-efficient of dissipation
($\eta$) is related simply to the normal state conductivities
of the superconducting sheet ($\sigma_S$) and the 
metallic sheet ($\sigma_E$) by the simple relation
$\eta \propto {\sigma_E \sigma_S\over \sigma_E+\sigma_S}$.
Thus the amount of dissipation can be tuned
by changing the conductivity in either sheet.

Finally, let us summarise the main results of this paper.
We have obtained the effective action for the phase
fluctuations in a two-dimensional superconductor in close proximity
to a metallic sheet. We find that the long wavelength
($q< d^{-1}$) modes are dissipative.
The coefficient of dissipation is related to the 
normal state conductivities of both layers.
The short wavelength ($q>d^{-1}$) modes are
unaffected by the presence of the metallic sheet.
The wavevector $d^{-1}$ defines a crossover scale
for the behaviour of the phase fluctuations
to change from one regime (dissipative)
to the other regime (propogating).

\noindent {\bf Acknowledgements}:- We thankfully acknowledge stimulating
 and
clarifying discussions with T. V. Ramakrishnan,  M. Randeria
and A. Taraphder.
\newpage
\noindent {\bf References}

\begin{enumerate}

\item S. Doniach, {\em Phys. Rev.} {\bf B24}, 5063 (1981).

\item S. Chakravarty, G-L. Ingold, S. Kivelson and G. Zimanyi,
{\em Phys. Rev.} {\bf B37}, 3283 (1988).

\item M. P. A. Fisher, {\em Phys. Rev.} {\bf B36}, 1917 (1987).

\item D. B. Haviland, Y. Liu and A. M. Goldman,
{\em Phys. Rev. Lett.} {\bf 62}, 2180 (1989).

\item L. J. Geerligs et. al., {\em Phys. Rev. Lett.} 
{\bf 63}, 326 (1989).

\item  H. Ding et. al., {\em Phys. Rev. Lett.}
 {\bf 78}, 2628 (1997).
 
 \item J. M. Harris et. al., {\em Phys. Rev. Lett.}
  {\bf 79}, 143 (1997).

\item Ch. Renner, B. Revaz, J-Y. Genoud, K. Kadowaki
and O. Fischer, {\em Phys. Rev. Lett.} {\bf 80}, 149 (1998).

\item V. J. Emery and S. A. Kivelson, {\em Nature} 
{\bf 374}, 434 (1995).

\item M. Franz and A. J. Millis, (unpublished).

\item A. J. Rimberg et. al. {\em Phys. Rev. Lett.} 
{\bf 78}, 2632 (1997).

\item U. Eckern, G. Schon and V. Ambegaokar,
{\em Phys. Rev.} {\bf B30}, 6419 (1984).

\item T. V. Ramakrishnan, {\em Physica Scripta}  {\bf T27}, 24 (1989).

\end{enumerate}

\end{document}